\documentclass[usenatbib]{mnras}
\usepackage{graphicx}	
\usepackage{multicol}	
\usepackage[english]{babel}
\usepackage{epstopdf}
\usepackage[T1]{fontenc}
\usepackage{ae,aecompl}
\usepackage{amsmath}    
\usepackage{amssymb}    

\def\degree{_\cdot^\circ}

\def\mpc{\,h^{-1}{\rm Mpc}}
\def\kpc{\,h^{-1}{\rm kpc}}

\def\msun{\,h^{-1}{\rm M}_\odot}

\def\subhalo{{\tt subhalo}}

\usepackage{color}

\makeatletter

\newcommand{\Rmnum}[1]{\expandafter\@slowromancap\romannumeral #1@}
\makeatother

\title[galaxy-halo alignments]{Galaxy-group (halo) alignments from SDSS DR7 and
the ELUCID simulation}

\author[Youcai Zhang]{Youcai Zhang$^{1}$\thanks{yczhang@shao.ac.cn},
Xiaohu Yang$^{2,3}$,
Hong Guo$^{1}$
\\
$^{1}${Key Laboratory for Research in Galaxies and Cosmology,
  Shanghai Astronomical Observatory; Nandan Road 80, Shanghai 200030,
  China} \\
$^{2}$ Department of Astronomy, School of Physics and
  Astronomy, Shanghai Jiao Tong University, Shanghai 200240, China \\
  $^{3}$ Tsung-Dao Lee Institute, and Shanghai Key Laboratory
  for Particle Physics and Cosmology, Shanghai Jiao Tong University, \\~~~~Shanghai 200240, China
}

\begin{document}
\label{firstpage}
\pagerange{\pageref{firstpage}--\pageref{lastpage}}
\maketitle

\begin{abstract}

Based on galaxies from the Sloan Digital Sky Survey (SDSS) and subhalos in the corresponding reconstructed region from the constrained simulation of ELUCID, we study the alignment of central galaxies relative to their host groups in the group catalog, as well as the alignment relative to the corresponding subhalos in the ELUCID simulation. Galaxies in observation are matched to dark matter subhalos in the ELUCID simulation using a novel neighborhood abundance matching method. In observation, the major axes of galaxies are found to be preferentially aligned to the major axes of their host groups. There is a color dependence of galaxy-group alignment that red centrals have a stronger alignment along the major axes of their host groups than blue centrals. Combining galaxies in observation and subhalos in the ELUCID simulation, we also find that central galaxies have their major axes to be aligned to the major axes of their corresponding subhalos
in the ELUCID simulation. We find that the galaxy-group and galaxy-subhalo alignment signals are stronger for galaxies in more massive halos. We find that the alignments between main 
subhalos and the SDSS matched subhalo systems in simulation are slightly stronger than the galaxy-group alignments in observation.

\end{abstract}


\section{Introduction}\label{sec_intro}

In the current paradigm of hierarchical structure formation, dark matter halos form through the gravitational collapse of the initial density perturbation and evolve via mass accretion and mergers with other halos \citep{White1978}. Since galaxies form and evolve in the potential well of dark matter halos, it is expected that the orientations of galaxies are related to those of their host halos, as well as the large-scale environment. Consequently, understanding the relation between the shapes of galaxies and halos is one of the most important parts of galaxy formation and evolution. Meanwhile, the alignment between galaxies and halos can be used to constrain the systematic errors in the weak gravitational lensing measurements \citep{Codis2015, Joa2015, Kilbinger2015, Kirk2015, Fortuna2020}.

In the past decades, numerous studies have characterized various types of alignments among galaxies, halos and the large-scale environment. The large-scale environment is commonly characterized by the tidal field to define the orientations of filaments or sheets in the cosmic web, leading to studies of the alignment of galaxies or halos with respect to their large-scale environment \citep{Zhang2009, Zhang2013, Zhang2015, Tempel2015, Hirv2017, Xia2017, Gane2018, Chen2019, Gane2019,Lee2019, Kraljic2020}. Observationally, the galaxy groups are usually adopted to study the alignment of the shapes of centrals, satellites or galaxy groups with respect to central-satellite position vector \citep{Brain2005, Pere2005, Agu2006, Yang2006, Fal2007, Kang2007, Siverd2009, Agu2010, Hao2011, LiZhigang2013, Chisari2014, Singh2015, Huang2018, Georgiou2019}, as well as the alignment between the major axes of the galaxies and those of their host groups \citep{WangY2008, Nie2010, Huang2016, West2017,Ume2018}. The observational evidence shows that the major axes of central galaxies are well aligned with their host clusters, and the alignment signals are stronger for redder and more luminous galaxies. 
For example, \citet{Huang2016} measured the projected alignments between the major axes of the central galaxies and those of their host clusters using three different
galaxy shape measurement methods. The average alignment angle based on isophote shape measurement is about $33$ degrees in the two-dimensional case.

Besides the few observational measurements using galaxy clusters, a vast majority of studies in literature rely on the cosmological hydrodynamical simulations to determine the alignment between galaxies and their host halos \citep{Bett2010, Okabe2018, Okabe2020, Brain2019, Bho2020, Soussana2020, Tang2020, Tenneti2020}. There is general agreement in literature that the major axes of central galaxies are well aligned with their halos, and the alignment signals are stronger galaxies in more massive halos. However, the average alignment angle measured in the simulations spans a wide range from $30^\circ$ to $50^\circ$ in the three-dimensional case \citep[see e.g.,][]{Tenneti2014,Vell2015, Shao2016, Chisari2017}. The wide range of variations on the alignment signals from cosmological simulations may be caused by the adoption of different hydro subgrid physics modeling, different estimators in the quantification of position angle, and different
numerical techniques with or without adaptive mesh refinement \citep{Springel2010}.

The above two methods both have pros and cons. In observation, the orientations of the galaxy clusters determined from the satellite galaxy distribution may not be the exactly same as the halos, which would affect the accuracy of the alignment angle. On the other hand, the alignment angle measured in the hydrodynamical simulations apparently strongly correlates with the complex baryonic physics in the galaxy formation and evolution models. For example, \citet{Tenneti2017} found that the orientation of the stellar distribution can be affected by the angular momentum of the galactic wind used in the hydrodynamical simulation. In this paper, we propose a different method of measuring the alignment angle by connecting the observed galaxies to the dark matter
halos in the constrained $N$-body simulation of ELUCID \citep{WangHY2014}, which reasonably reproduces the large-scale environment of the observed galaxies.   

The initial condition of ELUCID is extracted from the 
density field of galaxies in the SDSS DR7 Main Sample \citep{WangHY2014}. The mass and positions of the halos at $z\sim0$ in the ELUCID simulation are consistent with those in the galaxy group catalogs, especially for halos with mass larger than $10^{13}\msun$. Using a novel neighborhood abundance matching method, \citet{Yang2018} built up the connection between galaxies in SDSS DR7 to the subhalos in the corresponding observed region in the ELUCID simulation. Based on their galaxy-subhalo connection, we are able to make a one-to-one comparison between the projected major axes of galaxies and those of the subhalos.

The paper is organized as follows. In Section~\ref{sec_data}, we present the observational data from SDSS DR7 and the subhalo catalog from the ELUCID simulation. In Section~\ref{sec_method}, we describe the novel neighborhood abundance matching method which links galaxies in SDSS DR7 to dark matter subhalos in the ELUCID simulation and present the method to calculate the shapes of dark matter subhalos in simulation and groups in observation. In addition, we describe the methodology to quantify the various alignment signals. In Section~\ref{sec_result},  we present the alignments of galaxies with respect to their host groups in observation and their corresponding subhalos in simulation. In Section~\ref{sec_summary}, we summarize our main results.

\section{Data}\label{sec_data}

\subsection{Galaxies in the SDSS DR7}

Based on galaxy sample of the New York
University Value-Added Galaxy Catalog \citep[NYU-VAGC;][]{Blanton2005}, constructed from the SDSS DR7 \citep{Abazajian2009}, we collect a total of $639,359$ galaxies in the redshift range of $0.01 \le z \le 0.2$. We use $472,416$ groups identified from these galaxies with the adaptive halo-based group finder as in \citet{Yang2005} and \citet{Yang2007}. Most of groups only have a single member galaxy, and there are $68,170$ groups with at least two members. We refer the readers to \citet{Yang2007} for details. 

Following \citet{Yang2008}, we separate the galaxies into red and blue subsamples according to the $g-r$ color cut in the absolute magnitudes,
\begin{equation}
\centering
g-r = 1.022 - 0.0651x - 0.00311x^2,
\label{eqn:color}
\end{equation}
where $x=M_r - 5\log h + 23.0$ and $M_r$ is the absolute r-band magnitude k+e corrected to $z=0.1$. Since the initial condition of the ELUCID simulation is constructed from the distribution of the galaxies in the continuous Northern Galactic Cap (NGC), in this paper we only select galaxies in the NGC region of the range $99^{\circ} < \alpha < 283^\circ$ and $-7^{\circ}<\delta < 75^{\circ}$ , where $\alpha$ and $\delta$ are the right ascension and declination, respectively. This results in a final sample of $396,069$ galaxies in the NGC.

\subsection{Subhalos in the ELUCID simulation}

The halo catalog used in this study is from the ELUCID simulation  \citep{WangHuiyuan2016}, which is a dark matter only, constrained simulation carried out in the Center for High Performance Computing of Shanghai Jiao Tong University using L-GADGET2 code, a memory-optimized  version of GADGET2 \citep{Springel2005}. The simulation evolves $3072^3$ dark matter particles in a periodic box of $500\mpc$ on a side from redshift $z=100$ to the present epoch. The particle mass and softening length are $3.0875 \times 10^8 \msun$ and $3.5 \kpc$, respectively. The cosmological parameters adopted in the simulation are  $\Omega_{\rm m} = 0.258$, $\Omega_{\rm b} = 0.044$, $\Omega_\Lambda = 0.742$, $h =0.72$, $n_s = 0.963$ and $\sigma_8 = 0.796$, where $\Omega_{\rm m}$ is the matter density, $\Omega_{\rm b}$ is the baryon density, $\Omega_\Lambda$ is the dark energy density, $h$ is the normalized Hubble constant today, $n_s$ is the power law index of the initial power spectrum, and $\sigma_8$ is the amplitude of matter fluctuations on $8\mpc$ scales today.

Dark matter halos are identified using the standard friends-of-friends (FOF) algorithm \citep{Davis1985} with a linking length of $b=0.2$ times the mean particle separation and containing at least $20$ particles. Since the FOF algorithm may accidentally identify two independent structures linking with particle bridge as one structure, it is necessary to detect substructures inside a larger FOF halo. In this paper, a $\subhalo$ is defined as the locally overdense, self-bound substructure within a larger parent FOF halo. Using the algorithm SUBFIND designed by \citet{Springel2001}, we decomposed a given FOF halo into a set of disjoint self-bound subhalos. The center of the subhalo is defined as the position of the most bound particle, and the velocity is  defined as the mean velocities of the particles in the subhalo. Within these subhalos, the most massive one is regarded as the main subhalo. Obviously, compared with FOF halos, self-bound subhalos are more suitable to link central or satellite galaxies in observation.

\section{Methods}\label{sec_method}

\subsection{Matching between galaxies and subhalos}

Since the initial condition of the ELUCID simulation is constrained by the mass density field extracted from the galaxy distribution in observation, it is expected that the subhalo distribution at present is tightly correlated with the spatial distribution of galaxies in SDSS DR7.

\citet{Yang2018} proposed a novel neighborhood abundance matching method to link galaxies in SDSS DR7 to dark matter subhalos in the ELUCID simulation, according to the likelihood of the subhalo to be linked to the candidate galaxy \begin{equation}\label{eqn:match} P = M_{\rm sh} \exp \left( - \frac {r_p^2} {2 r_{\rm off}^2} \right) \exp \left( -\frac {\pi^2} {2 v_{\rm off}^2}  \right), \end{equation} where $r_p$ and $\pi$ are the separations between the galaxy and the subhalo in the perpendicular and parallel to the line-of-sight direction, $M_{\rm sh}$ is the mass of the subhalo under consideration, $r_{\rm off}$ and $v_{\rm off}$ are two free parameters. Note that $r_{\rm off} = \infty$ and $v_{\rm off} = \infty$ correspond to the traditional abundance matching method. Here, the parameters are set to be $r_{\rm off} = 5 \mpc$ and $v_{\rm off} = 1000 ~{\rm km/s}$.  This results in a total of $396,069$ galaxy-subhalo pairs in the catalog.

In what follows, we use the galaxy-subhalo pairs by matching separately for central and satellite galaxies. Central galaxies are linked to the main subhalos in the host FOF halos and satellite galaxies to the other subhalos. The matching criteria can give better constraint for the luminosity (stellar mass)-subhalo mass relation. This results in a total of $396,069$ galaxy-subhalo pairs in the matching catalog. Using the galaxy-subhalo pairs, \citet{Yang2018} reproduced the satellite fraction, the conditional luminosity function(the conditional stellar mass function), and the biases of the galaxies.

In this paper, only galaxies in groups with at least two member galaxies are selected to measure the alignment signals. After this selection, there are $43,316$ central galaxies and $118,095$ satellite galaxies.

\subsection{Shape and alignment definition}

The shape of a subhalo containing $N$ dark matter particles is calculated by the simple inertia tensor \begin{equation}\label{eqn:it} I_{\alpha \beta} = m  \sum\limits_{i=1}^N x_{i,\alpha} x_{i,\beta}, \end{equation} where $m$ is the particle mass, $\alpha$ and $\beta$ are the inertia tensor indices with values of $1$, $2$ or $3$, and $x_{i,\alpha}$ is the position of the particle $i$ with respect to the center of the subhalo, which is defined as the position of the particle with the minimum potential. The axis lengths $a$, $b$ and $c$ ($a\geq b \geq c$) of the ellipsoidal subhalo are proportional to the square roots of the eigenvalues $\lambda_1$, $\lambda_2$ and $\lambda_3$ ($\lambda_1 \geq \lambda_2 \geq \lambda_3$) of the inertia tensor by $a=\sqrt{\lambda_1}$, $b=\sqrt{\lambda_2}$ and $c=\sqrt{\lambda_3}$ . The corresponding eigenvectors denote the directions of the major, middle and minor axes of the subhalos.

In order to calculate the projected orientation on the sky of the halos, the Cartesian coordinates in the simulation are transformed into the redshift $z$ and sky coordinates $\alpha$ and $\delta$, where $\alpha$ and $\delta$ are the right ascension and declination, respectively. For a subhalo at the location $\boldsymbol x$ with the three-dimensional direction of the major axis $\Delta{\boldsymbol x}$, the projected direction $\theta_{\rm H}$ on the sky can be calculated by \begin{equation}\label{eqn:project} \theta_{\rm H} = \arctan \left( \frac {\Delta \alpha \cos \delta} {\Delta \delta} \right), \end{equation} where $\Delta\alpha$ and $\Delta\delta$ are the right ascension and declination differences between the locations $\boldsymbol x$ and $\Delta \boldsymbol x$.

In observation, for each group with more than one member, the group shape is calculated using the projected two-dimensional case of Equation~\ref{eqn:it}, where $N$ is the number of the galaxies in the group, $x_{i,1} = \Delta \alpha \cos \delta$ and  $x_{i,2}= \Delta \delta $ are the $i$-th projected coordinates on the sky with the origin at the position of the central galaxy, where $\Delta\alpha$ and $\Delta\delta$ are the right ascension and declination differences between the locations of satellite galaxies and central galaxies. For each galaxy, the orientation angle $\theta_{\rm G}$ of the major axis of the galaxy on the sky is specified by the $25$ mag~arcsec$^{-2}$ isophote in the r-band.

In order to quantify the alignment signal between  the orientation of the major axis of a galaxy in observation and its corresponding subhalo in the ELUCID simulation, we calculate the normalized probability distribution of the angle $\theta$ between the two orientations as
\begin{equation}
P(\theta) = N(\theta)/\langle N_{\rm R}(\theta)\rangle,
\end{equation}
where $N(\theta)$ is the number of the galaxy-subhalo or galaxy-group pairs in each $\theta$ bin, and $\langle N_R(\theta)\rangle$ is the average number of such pairs obtained from $100$ random samples, in which the orientation of galaxies are kept fixed, but the other orientations (subhalos or groups) are randomized. The standard deviation of $P_{\rm R}(\theta) = N_{\rm R}(\theta)/\langle N_R(\theta)\rangle$ calculated from the random samples is used to assess the  significance of the deviation of $P(\theta)$ from unity. We note that $P(\theta)=1$ is the case without any alignment. Since the significance is quantified with respect to the null hypothesis, throughout the paper the error bars are plotted on top of $P(\theta)=1$ line. The angle $\theta$ is constrained in the range of $0^{\circ}\leq \theta \leq 90^{\circ}$. For two parallel orientations, $\theta = 0$, while for two perpendicular orientations, $\theta = 90$. In addition, we calculate the average angle $\langle \theta \rangle$ and $\theta_\sigma$, which is the standard deviation of $\langle \theta_{\rm R} \rangle$ of the $100$ random samples. In the absence of the alignment, $\langle \theta \rangle = 45^{\circ}$, however $\langle \theta \rangle = 45^{\circ}$ does not necessarily mean a isotropic distribution.

\section{Results}\label{sec_result}

\subsection{galaxy-group alignment in observation}

\begin{figure}
\includegraphics[width=0.5\textwidth]{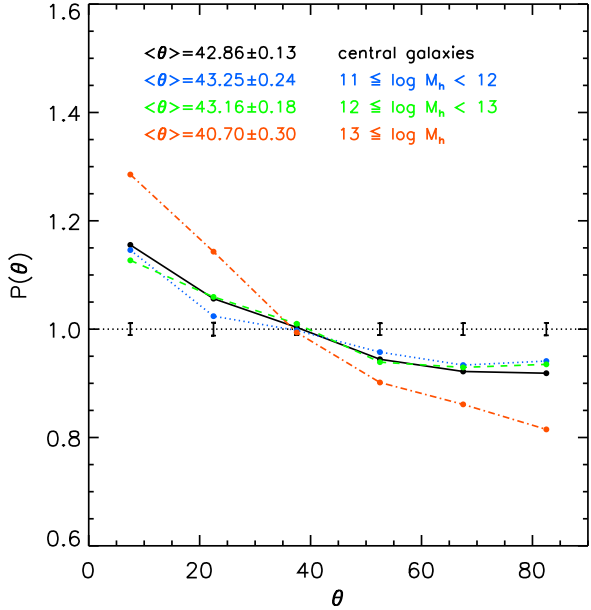}
\caption{The normalized probability distribution of the angles between
the major axes of central galaxies and the major axes of their host groups
in observation. The black solid line shows the alignment signal for a
total of $43,316$ central galaxies in the groups with member galaxies
larger than $2$. The horizontal line shows an isotropic distribution of
alignment angles, while the error bars are obtained from $100$ random
realizations. The average angle and its error are also indicated.}
\label{fig:gg_cen}
\end{figure}

\begin{figure}
\includegraphics[width=0.5\textwidth]{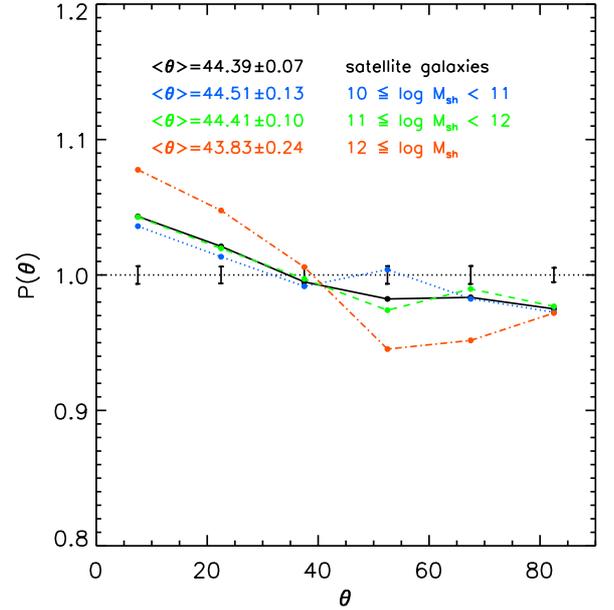}
\caption{Similar to Figure~\ref{fig:gg_cen}, but for the alignments
between satellite galaxies and their host groups. }
\label{fig:gg_sat}
\end{figure}

\begin{figure}
\includegraphics[width=0.5\textwidth]{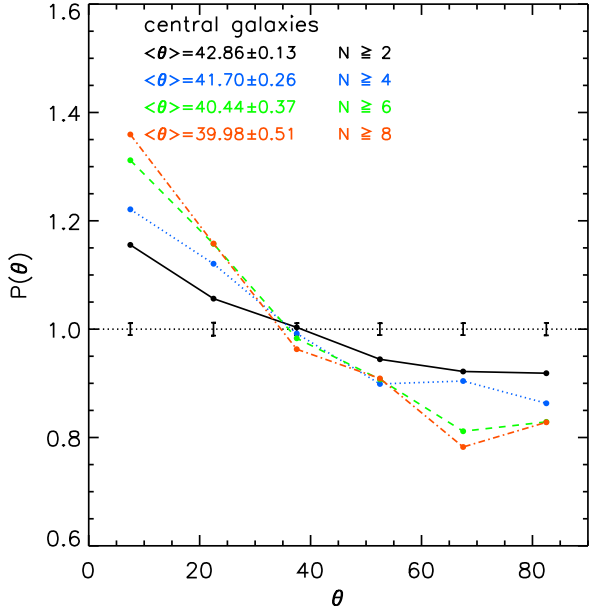}
\caption{Galaxy-group alignment for central galaxies in groups with
member galaxies $N \geq 2$, $N \geq 4$, $N \geq 4$, and $N \geq 8$,
respectively. }
\label{fig:gg_member}
\end{figure}

\begin{figure}
\includegraphics[width=0.5\textwidth]{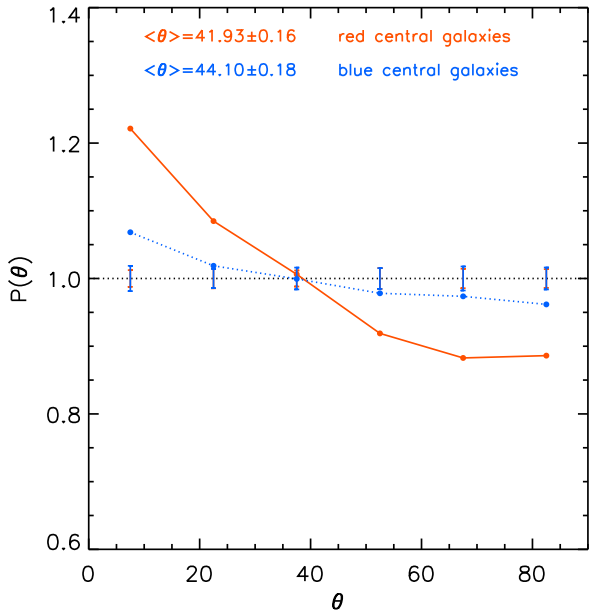}
\caption{Galaxy-group alignment for red and blue central galaxies. The 
error bars on top of the horizontal line are also shown for red and
blue subsamples, respectively. }
\label{fig:gg_rb}
\end{figure}

In this subsection, we measure the alignment signal between the major axes of the galaxies in the SDSS DR7 group catalog and those of their host groups. For each group with at least two members, we calculate the shape of the group using the projected two-dimensional case of Equation~\ref{eqn:it} based on the angular positions of satellite galaxies with respect to their central galaxies. There are a total of $43,316$ groups with more than one member for $396,069$ galaxies in the continuous region.  Figure~\ref{fig:gg_cen} shows the probability distribution $P(\theta)$, with the black line for all centrals. With an average alignment angle of $\langle \theta \rangle = 42\degree86\pm 0\degree13$ for a total of $43,316$ central galaxies, there is a clear trend that the major axes of central galaxies are preferentially aligned parallel to the major axes of their host groups. The error bars of the measurements are shown on top of the horizontal dotted line taken from $100$ realizations, in which the major axes of the groups have been randomized. 

In observation, based on the redMapper cluster catalog from SDSS DR8, \citet{Huang2016} also found that the shapes of central galaxies are aligned with the shapes of their parent clusters traced by the satellite location in the clusters at redshifts $0.1<z<0.35$. Using a sample of $65$ distant galaxy clusters from HST observation at redshifts $0.19<z<1.8$, \citet{West2017} reported the similar alignment signals, which are also confirmed in cosmological hydrodynamical simulations \citep{Shao2016,Tenneti2020}.

To study the mass dependence of the alignment signal, the central galaxies in the galaxy-group pairs are then separated into three different halo mass subsamples with $\log (M_{\rm h}/\msun)$ in the ranges of $(11,12)$, $(12,13)$, and $(13,\infty)$. Here for each central galaxy, the halo mass is taken from its matched main subhalo in the ELUCID simulation. The alignment signals for the subsamples are shown using different types of lines with different colors in Figure~\ref{fig:gg_cen}. There is a clear indication that the alignment is stronger for more massive halos. Similar trend are also found for the galaxy stellar mass.

We also study the alignment signals of satellite galaxies with respect to their host groups, as shown in Figure~\ref{fig:gg_sat}. Similar trends of alignment angle distribution, as well as the halo mass dependence, are found for the satellite galaxies. The alignment signals of satellite galaxies with $\langle \theta \rangle = 44\degree39\pm 0\degree07$ are weaker than those of central galaxies with $\langle \theta \rangle = 42\degree86 \pm 0\degree13$. In the following analysis, we mainly focus on the  central galaxies.

To investigate the effect of group richness, all groups are separated into different richness subsamples, as shown in Figure~\ref{fig:gg_member}. Central galaxies in richer groups have higher chance to be aligned with their host groups, confirming the result of \citet{Nie2010} using the galaxy cluster catalog from SDSS DR6.

Finally, we show the dependence of the alignment signal on the galaxy color in Figure~\ref{fig:gg_rb}, where we separate the galaxies into red and blue subsamples according to Equation~(\ref{eqn:color}). The alignment signal for the red centrals with $\langle \theta \rangle = 41\degree93\pm 0\degree16$ is much stronger than that of the blue centrals with $\langle \theta \rangle = 44\degree10\pm 0\degree18$, similar to the finding of \citet{Agu2010}.

\subsection{Alignment between galaxies in observation and subhalos in simulation}

\begin{figure}
\includegraphics[width=0.5\textwidth]{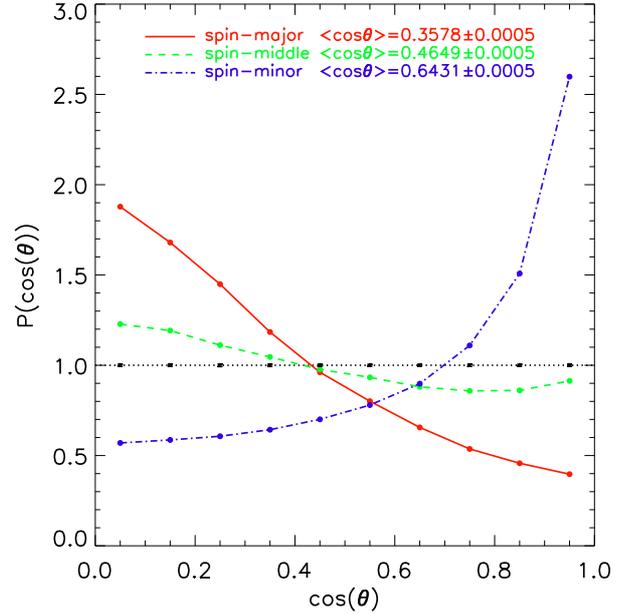}
\caption{The probability distribution of the angle $\cos\theta$
between the angular momentum and the  major (red),
middle (green), and minor (blue) axes of the subhalos in 
the ELUCID simulation in three-dimensional case.}
\label{fig:ss_3d}
\end{figure}

\begin{figure}
\includegraphics[width=0.5\textwidth]{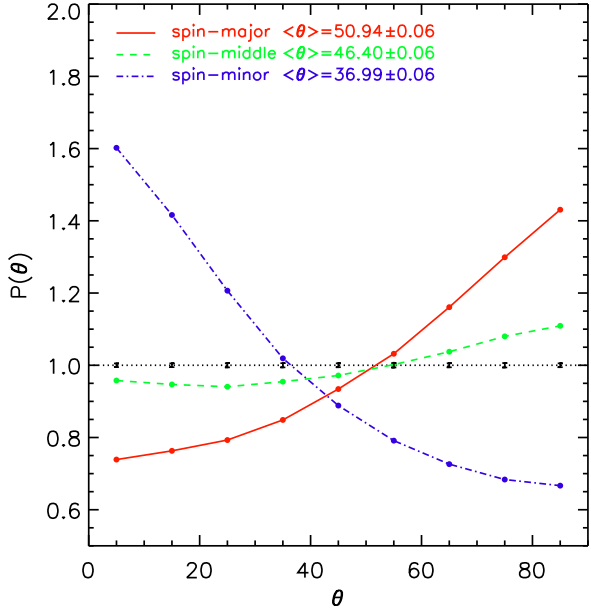}
\caption{The probability distribution of the angle $\theta$
between the projected angular momentum and the projected major (red),
middle (green), and minor (blue) axes in two-dimensional case.}
\label{fig:ss_2d}
\end{figure}

\begin{figure}
\includegraphics[width=0.5\textwidth]{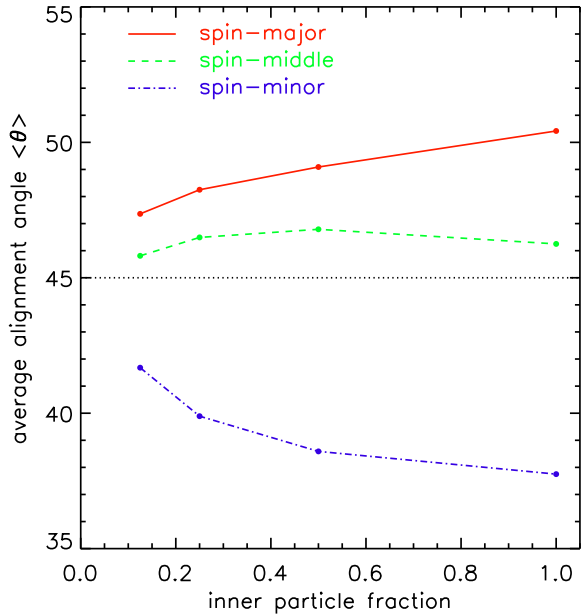}
\caption{The average alignment angle $\langle \theta \rangle$
as a function of the inner particle fraction of the subhalos in  the
ELUCID simulation for spin-major (red), spin-middle (green), and 
spin-minor (blue) alignments.}
\label{fig:sa}
\end{figure}

In this section, we investigate the alignment signal between the major axes of the SDSS galaxies and those of matching subhalos in the ELUCID simulation \citep{Yang2018}. As the satellite galaxy-group alignment is very week, in what follows, we only focus on the central-main subhalo pairs.

\subsubsection{spin-shape alignment in simulation}

In order to check the reliability of the shape calculation for subhalos in simulation, we first examine the spin-shape alignment within the subhalos themselves in the ELUCID simulation. The subhalo spin vector is defined as \begin{equation}\label{eqn:spin} \boldsymbol J = m \sum \limits_{i=1}^N \boldsymbol r_i \times \boldsymbol v_i, \end{equation} where $m$ is the dark matter particle mass, $\boldsymbol r_i$ is the position vector of the $i$-th particle with respect to the subhalo center of mass, and $\boldsymbol v_i$ is the velocity of the $i$-th particle relative to the bulk velocity of the subhalo. 

Previous studies mainly focus on the spin-shape alignment in the three-dimensional case. For comparison,  Figure~\ref{fig:ss_3d} shows probability distribution of $\cos\theta$, where $\theta$ is the three-dimensional alignment angle between the major axes of subhalo shape and the subhalo spin vector. Note that two parallel orientations correspond to $\cos\theta = 1$, while two perpendicular orientations correspond to $\cos\theta =0$. As shown in Figure~\ref{fig:ss_3d}, the spin direction is preferentially parallel to the subhalo minor axis, which is in good agreement with the previous results \citep{Bett2007, Zhang2009, Chisari2017, Gane2018}.

The spin direction of the subhalo in three-dimensional case is calculated using Equation~\ref{eqn:spin}, and then projected on the sky using Equation~\ref{eqn:project}, similar to the projected shape orientations \citep{Zhang2013, Zhang2015}. Figure~\ref{fig:ss_2d} shows the alignment signals of Figure~\ref{fig:ss_3d} in the projected two-dimensional case. The spin direction has a strong tendency to be parallel to the minor axis with $\langle \theta \rangle = 36\degree99\pm 0\degree06$, and much weaker chance to be perpendicular to the major axes with $\langle \theta \rangle = 50\degree94\pm 0\degree06$.

The spin and shape orientations have been proven to vary as a function of distance from the center \citep{Bailin2005, Bett2010}. For each subhalo, we also calculate the spin and shape orientations for the inner subhalo regions consisting of the inner $1/8$, $1/4$, and $1/2$ of the subhalo particles. Figure~\ref{fig:sa} shows the average angle $\langle \theta \rangle$ of the spin-shape alignment in two-dimensional case as a function of the inner particle fraction of the subhalos in the ELUCID simulation.  As indicated in Figure~\ref{fig:sa}, the spin-shape alignment is dependent on the radial extent of the subhalo. Both the spin-minor and spin-major alignment are weaker in the inner part, compared to those of the entire subhalos. In the following analysis, we will mainly use the subhalo shape calculated from the entire subhalo particles.

\subsubsection{galaxy-subhalo alignment}

\citet{WangHY2012} generated the initial condition of the constrained ELUCID simulation based on the galaxy density field using the groups with halo masses larger than $10^{12}\msun$. \citet{Tweed2017} compared the halos in reconstructed and original simulations and found that the reconstruction techniques are reliable for the most massive halos with masses larger than $5\times10^{13} \msun$, where more than half of the halo particles in the original simulation are matched in the reconstructed one.

In the neighborhood abundance matching method developed by \citet{Yang2018}, for each galaxy, the likelihood of the corresponding subhalo calculated by Equation~\ref{eqn:match} is proportional to the subhalo mass in a certain small volume. In this method, \citet{Yang2018} started from the most massive galaxy to search for its corresponding subhalo. Therefore, the most massive subhalo in the searching volume is mostly linked to the most massive galaxy, leading to the fact that matching pairs of central galaxies are more reliable than those of satellite galaxies. In the following analysis, we mainly focus on the alignment signals of central galaxies.  In total, there are $43,316$ central galaxies in groups with at least two members.

\begin{figure}
\includegraphics[width=0.5\textwidth]{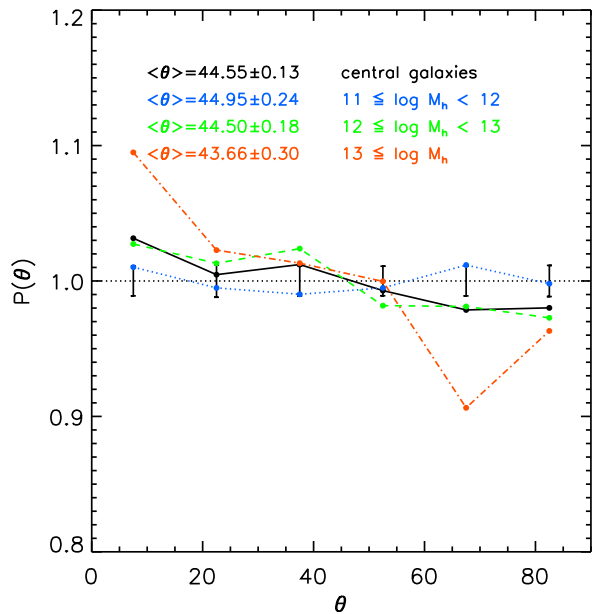}
\caption{The probability distribution of the angles between the major 
axes of the galaxies in observation and the projected major axes of 
their embedding main subhalos from the ELUCID simulation. 
The black solid 
line shows the alignment signal for a total of $43,316$ central galaxies
in the galaxy-halo
pairs catalog. The horizontal line shows an isotropic distribution of
alignment angles, while the error bars are obtained from $100$ random
realizations for total galaxies. The average angle and its error are 
also indicated.}
\label{fig:gh}
\end{figure}

Figure~\ref{fig:gh} shows the probability distribution of the angles between the major axes of the central galaxies from the SDSS DR7 and the projected major axes of their host subhalos in the ELUCID simulation. With an average alignment angle of $\langle \theta \rangle = 44\degree55 \pm 0\degree13 $ for $43,316$ central galaxies, there is a clear tendency that the major axes of central galaxies are preferentially parallel to those of their host subhalos. However, the alignment signals are weaker than those of galaxy-group alignment. Here the shapes of the subhalos in simulation are computed from the distribution of the entire dark matter particles in the subhalos. We have checked the alignment signals only using the inner $1/8$ particles in the subhalos, and the alignment signals are still weak between galaxies and the inner region of the subhalos. 

Besides, we have checked the alignment signals using the galaxy-subhalo pairs with smaller separation $r_p$ and $\pi$ in Equation~\ref{eqn:match}. The alignment signals are slightly
stronger for the galaxy-subhalo pairs with smaller separation.  For $50\%$ cleaner galaxy-subhalo pairs with $r_p < 2.8 \mpc$ and $\pi < 3.2 \mpc$, the average angle of 
galaxy-subhalo alignment is $44\degree15 \pm 0\degree25$.

We also study the mass dependence of the alignment signals using the subsamples separated by the embedding subhalo mass with $\log (M_{\rm h} /\msun)$ in the ranges of $(11,12)$, $(12,13)$, and $(13,\infty)$. In Figure~\ref{fig:gh}, different types of lines with blue, green and red colors show the alignment signals for galaxies in different subhalo mass ranges. There is a clear indication that galaxies in more massive subhalos have stronger alignment signals. Based on the previous studies \citep{WangHY2012, Tweed2017, Yang2018}, the cross-identification method is more reliable for more massive subhalos. Besides, the shape measurements are more accurate for subhalos containing more dark matter particles. Obviously, the results of galaxies in subhalos with $\log (M_{\rm h}/ \msun) \geq 13 $ are more reliable. As shown by the red line in Figure~\ref{fig:gh}, there is a clear and significant alignment signal with $\langle \theta \rangle = 43\degree66\pm 0\degree30$ for galaxies in subhalos of $\log (M_{\rm h}/ \msun) \geq 13$.

\subsection{The impact of survey selection effects on the alignment signals}

\begin{figure}
\includegraphics[width=0.5\textwidth]{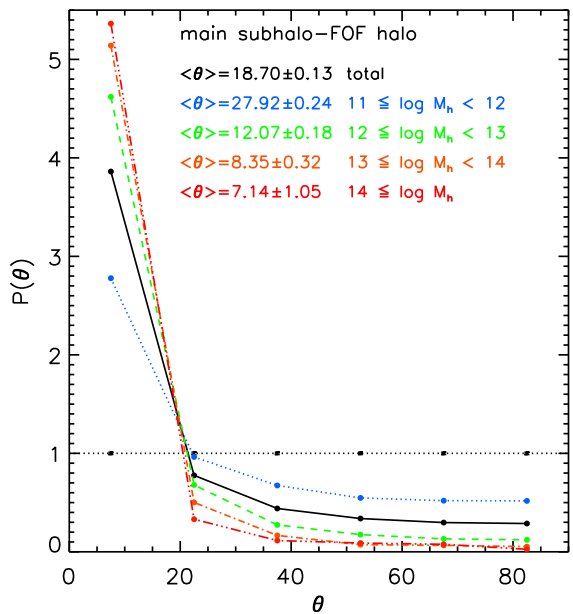}
\caption {The probability distribution of the angles between the projected major axes of the main subhalos and the projected major axes of FOF halos in simulation. 
The black solid line shows the result for a total of $43,316$ FOF halos. Different colors correspond to the subsamples with different main subhalo mass.}\label{fig:sub_fof_dm}
\end{figure}

There are a total of $43,316$ FOF halos in the ELUCID simulation corresponding to the observed groups with at least two members.
In this section, based on the subhalo and FOF halo catalogs from the ELUCID simulation, we investigate three types of alignment signals, which include the alignment 
between main subhalo shapes and the host FOF halos shapes traced by all dark matter particles, the alignment between main subhalo shapes and the satellite subhalo distribution systems,
and the alignment between main subhalo shapes and the SDSS matched satellite subhalo distribution systems. 

\subsubsection{Signals for dark matter particles }

Based on the subhalo and FOF halo catalogs from the ELUCID simulation, we first calculate the alignments between the main subhalos and their parent FOF halos, 
where the shapes of the FOF halos are calculated by the position distribution of all the dark matter particles in the FOF halos. The sample is separated into four subsamples according to 
the main subhalo mass, resulting in $15,288$, $19,379$, $5,765$, and $471$ subhalos with the mass $\log (M_{\rm h} /\msun)$ in the ranges of $(11,12)$, $(12,13)$, $(13,14)$ and $(14,\infty)$.

Figure~\ref{fig:sub_fof_dm} shows the alignments of the major axes of the main subhalos with respect to the FOF halo shapes traced by dark matter particles. With the average alignment angle $\langle \theta \rangle = 18\degree7\pm 0\degree13$, in dark matter only simulation, the major axes of the main subhalos are strongly correlated with the major axes of their parent
FOF halos. Besides, there is a mass dependence that the alignment signals are stronger for more massive halos.

\begin{figure}
\includegraphics[width=0.5\textwidth]{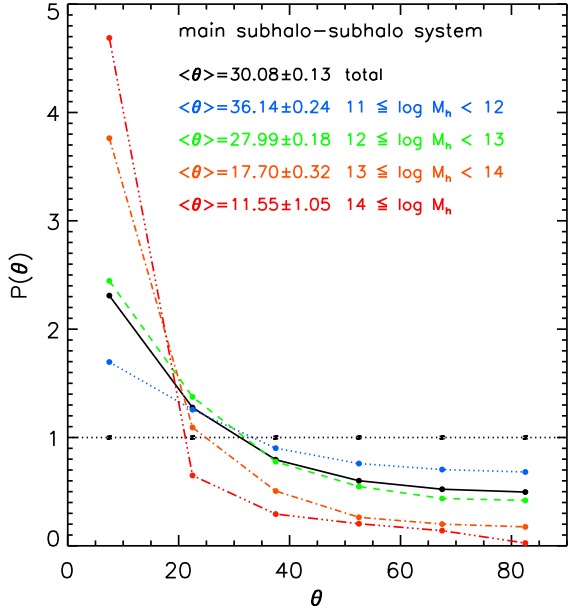}
\caption{The probability distribution of the angles between the projected major axes of the main subhalos and the distribution of all the subhalos in FOF halos in simulation. The black solid line shows the result for a total of $43,316$ halos. Different colors correspond to the subsamples with different halo mass.}\label{fig:sub_fof}
\end{figure}

\subsubsection{Signals for all the subhalos }

To fully understand the alignment signals in observation and the ELUCID simulation, we then calculate the alignments between the main subhalos and their subhalo systems, where the 
shapes of the subhalo systems are calculated by the position distribution of all the subhalos in the FOF halos.

Figure~\ref{fig:sub_fof} shows the alignments of the major axes of the main subhalos with respect to the distribution of all the subhalos in the FOF halos. There is a mass dependence that the alignment signals are stronger for more massive halos. For halo mass larger than $10^{14}\msun$, the average alignment angle is $\langle \theta \rangle = 11\degree55\pm 1\degree05$.

Such kind of mass dependence of the alignment signal is also found in the hydrodynamical simulations \citep{Vell2015, Tenneti2020}. From the Horizon-AGN simulation,  \citet{Okabe2020} claimed that the major axes of central galaxies are aligned with their subhalo systems with the average angle $\sim 20$ degrees in the projected plane for $40$ cluster-sized halos with masses larger that $5\times 10^{13}\msun$. As shown in Figure~\ref{fig:sub_fof}, the two-dimensional average alignment angle is $\langle \theta \rangle = 17\degree70\pm 0\degree32$ for halo masses of $ 13\leq \log (M_{\rm h}/\msun)\leq 14$, which are in agreement with the results of \citet{Okabe2020}. This agreement is expected in the fact that the galaxy-halo alignment in hydrodynamical simulation is determined, to first order, by the alignment between the inner dark matter distribution and the entire dark matter in the FOF halos. \citet{Vell2015} has claimed that, for stars and dark matter enclosed in sphere of the same radius, the orientation of the stellar distribution follows that of dark matter (see the right panel of their Figure~$8$), while dark matter itself changes orientation with increasing radii, resulting in the misalignment between galaxies and halos in hydrodynamical simulations.

\begin{figure}
\includegraphics[width=0.5\textwidth]{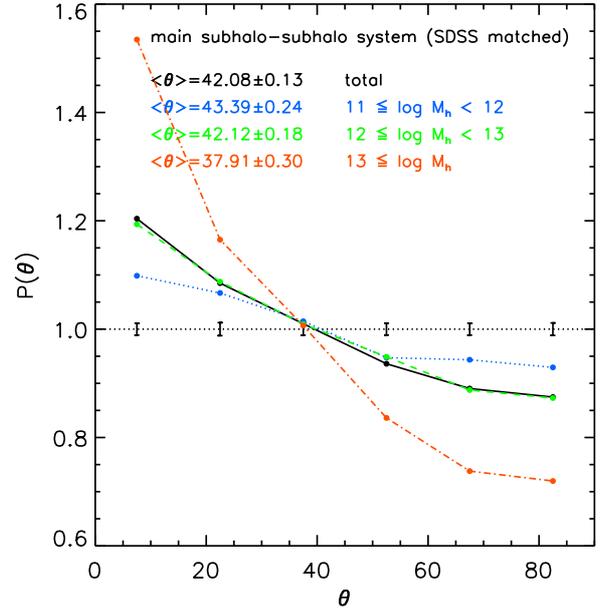}
\caption{The probability distribution of the angles between the projected major axes of the main subhalos and their hosts calculated using the subhalos linked with satellites 
in the groups from SDSS DR7.}\label{fig:sh_cen}
\end{figure}

\begin{figure}
\includegraphics[width=0.5\textwidth]{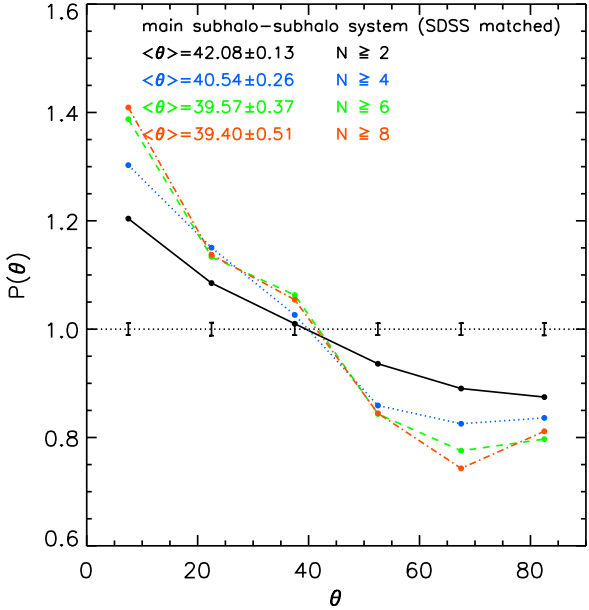}
\caption{Similar to Figure~\ref{fig:sh_cen}, but for the host system richness $N \geq 2$, $N \geq 4$, $N \geq 4$, and $N \geq 8$, respectively. }\label{fig:sh_member}
\end{figure}

\subsubsection{Signal for the observed subhalos}

In simulation, the alignments between main subhalos and their host halos are commonly much stronger than the alignments between galaxies and their host groups in observation, which may be due to the fact that the satellites subhalos resolved in N-body or hydrodynamical simulations are much more complete than the observation. To explore the impact of the survey selection effects, we use only the subhalos that are associated with the satellite galaxies in SDSS observation to measure the shape and orientation of subhalo systems.

In observation, the shape of the group is calculated by the distribution of the satellites in the group. For consistency, here the shape of the subhalo system in simulation is also calculated using the position distribution of the subhalos that are linked with the satellites in the observed group, which makes sure that the shape calculation of the subhalo system in simulation has used the same number $N$ in Equation~\ref{eqn:it} as that of the group in observation.

Figure~\ref{fig:sh_cen} shows the distribution of the angles between the projected major axes of 
the main subhalos and their subhalo systems, which are calculated only using the subhalos linked with 
satellites in the observed group. 
Similar to the galaxy-group alignments, there is a clear indication that the major axes of the main subhalos are preferentially aligned to the major axes of their subhalo systems in simulation. The alignments between the main subhalos and the SDSS matched satellite subhalo systems in
simulation with $\langle \theta \rangle = 42\degree08\pm 0\degree13$ are slightly larger than the galaxy-group alignments in observation with 
$\langle \theta \rangle = 42\degree86\pm 0\degree13$. As shown by the red lines in Figure~\ref{fig:sh_cen},
there is a clearly significant alignment signal with $\langle \theta \rangle = 37\degree 91\pm 0\degree30$ for $6,236$ halos with mass larger than $10^{13}\msun$, again somewhat larger than the signals obtained from the observations.

We also investigate the dependence of the alignment signals on the richness. The host systems are separated into different subsamples according to the richness at least $2$, $4$, $6$, and $8$. Figure~\ref{fig:sh_member} shows the alignment signal as a function of the richness. Similar to the tendency of the galaxy-group alignment, the alignment signals are stronger in richer systems. For host systems with the richness $N \geq 8$,  there is a clearly significant alignment with $\langle \theta \rangle = 39\degree40\pm 0\degree51$, which is in good agreement with the galaxy-group alignment with $\langle \theta \rangle = 39\degree98\pm 0\degree51$ with $N \geq 8$.

The results demonstrate that using only a small fraction of galaxies to trace the shape of host halo will significantly reduce the alignment signals between the orientations of the main subhalo and FOF host halo. Given that the ELUCID simulation can only reproduce the most massive halos, this results in a further 
reduction of the alignment signals.

Taken the results shown in Figure~\ref{fig:sh_cen} as a benchmark for the case of central (main subhalo) - group shape (as traced by subhalos) alignment in the situation without baryonic effect from a dark matter only simulation, the lower alignment signals shown in Figure 1 highlight the importance of taking into account the baryon effect on the intrinsic alignment signals, as discussed in the works of \citet{Tenneti2017} and \citet{Soussana2020}.

\section{Summary}\label{sec_summary}

In this paper, we have studied the orientation of galaxies relative to their host groups in observation and their corresponding subhalos in the ELUCID simulation.

Observationally, from the group catalog, we select $43,316$ groups with member galaxies larger than $2$.  The orientations of the galaxies are defined by the position angles of the major axes specified by the $25$ mag~arcsec$^{-2}$ isophote in the r-band. The shapes of the groups are chracterized by the inertia tensor of member galaxies in the groups. Based on the $43,316$ groups from SDSS DR7, we have investigated the alignment between the major axes of galaxies and their host groups. We have found that the major axes of central and satellite galaxies have a tendency to be aligned with the major axes of their host  groups. The galaxy-group  alignment signals of satellite galaxies are weaker than those of central galaxies. There is a mass and richness dependence that the alignment signals are stronger for galaxies in massive halos and in richer groups.   In addition, the galaxy-group alignment is found to depend on the galaxy colors. Red centrals show a stronger alignment than that of blue centrals.

For 43,316 central galaxies in groups with member galaxies larger than $2$, we have matched 43,316 main subhalos from the ELUCID simulation using a novel neighborhood abundance matching method \citep{Yang2018}. From the ELUCID simulation,  we have calculated the shapes of the halos by the inertia tensor of thousands of dark matter particles within the halos.  The shapes of the halos are then projected on the sky using Equation~\ref{eqn:project}, in order to compare with the shapes of the galaxies in observation. Using 43,316 main subhalos matched to central galaxies in observation, we have examined the alignment between the major axes of galaxies and their corresponding subhalos. We find that central galaxies are  preferentially parallel to the major axes of their corresponding subhalos. Galaxies in more massive subhalos have stronger galaxy-subhalo alignment signals.

For $43,316$ main subhalos matched to central galaxies in observation, we have calculated the alignments between main subhalos and their host systems in simulation. The shapes of the host systems in simulation are calculated using the positions of the subhalos matched to galaxies in observation. Totally, the alignments between main subhalos and SDSS matched subhalo systems 
in simulation are slightly stronger than galaxy-group alignments in observation. Similar to the galaxy-group alignments in observation, the projected major axes of the main subhalos are 
aligned to the major axes of their subhalo systems in simulation.

Totally, the major axes of central galaxies, groups and halos are preferentially parallel each other. The alignment signals are stronger for galaxies in more massive halos.  Especially for $6,236$ central galaxies with their corresponding halo mass larger than $10^{13}\msun$, there are clearly significant alignment signals with the average angles $\langle \theta \rangle = 40\degree70\pm 0\degree30$, $\langle \theta \rangle = 43\degree66\pm 0\degree30$, and $\langle \theta \rangle = 37\degree91\pm 0\degree30$  of the galaxy-group, galaxy-subhalo, and main subhalo-subhalo system alignments, respectively.

In addition, we have examined the spin-shape alignments within the subhalos themselves in the ELUCID simulation. The spin vectors of the subhalos are calculated by the angular momenta of the subhalos. The major (minor) axes of the subhalos are preferentially perpendicular (parallel) to the directions of angular momenta, which is in good agreement with the previous studies \citep{Bett2007, Zhang2009, Chisari2017, Gane2018}.  For each subhalo, we also calculate the spin-shape alignment within the inner regions consisting of the inner $1/8$, $1/4$, and $1/2$ of the subhalo particles. The spin-shape alignment signals are found to be weaker in the inner part of the subhalos.

\section*{Acknowledgements}

We thank the anonymous referee for the helpful comments that significantly improve the presentation of this paper.
This work is supported by the national science foundation of China 
(Nos. 11833005,  11890692, 11621303), 111 project No. B20019 and
Shanghai Natural Science Foundation, grant No. 15ZR1446700.
We also thank the support of the Key Laboratory for Particle
Physics, Astrophysics and Cosmology, Ministry of Education.

This work is also supported by the High Performance Computing Resource
in the Core Facility for Advanced Research Computing at Shanghai
Astronomical Observatory.

Funding for the Sloan Digital Sky Survey IV has been provided by the
Alfred P. Sloan Foundation, the U.S. Department of Energy Office of
Science, and the Participating Institutions. SDSS acknowledges support
and resources from the Center for High-Performance Computing at the
University of Utah. The SDSS web site is www.sdss.org.

SDSS is managed by the Astrophysical Research Consortium for the
Participating Institutions of the SDSS Collaboration including the
Brazilian Participation Group, the Carnegie Institution for Science,
Carnegie Mellon University, the Chilean Participation Group, the
French Participation Group, Harvard-Smithsonian Center for
Astrophysics, Instituto de Astrof{\'i}sica de Canarias, The Johns
Hopkins University, Kavli Institute for the Physics and Mathematics of
the Universe (IPMU) / University of Tokyo, Lawrence Berkeley National
Laboratory, Leibniz Institut f{\"u}r Astrophysik Potsdam (AIP),
Max-Planck-Institut f{\"u}r Astronomie (MPIA Heidelberg),
Max-Planck-Institut f{\"u}r Astrophysik (MPA Garching),
Max-Planck-Institut f{\"u}r Extraterrestrische Physik (MPE), National
Astronomical Observatories of China, New Mexico State University, New
York University, University of Notre Dame, Observat{\'o}rio Nacional /
MCTI, The Ohio State University, Pennsylvania State University,
Shanghai Astronomical Observatory, United Kingdom Participation Group,
Universidad Nacional Aut{\'o}noma de M{\'e}xico, University of
Arizona, University of Colorado Boulder, University of Oxford,
University of Portsmouth, University of Utah, University of Virginia,
University of Washington, University of Wisconsin, Vanderbilt
University, and Yale University.

\section*{Data availability}
The data underlying this article will be shared on reasonable request to the corresponding author.

\bibliographystyle{mnras}
\bibliography{bibliography}

\bsp    
\label{lastpage}
\end{document}